\documentclass{article}

\usepackage{graphicx}
\usepackage[]{epsfig}
\usepackage{epstopdf}
\input{epsf.sty}
\usepackage{texdraw}
\newcommand {\bi} {\bibitem}
\newcommand {\be} {\begin{equation}}
\newcommand {\ee} {\end{equation}}
\newcommand {\bea} {\begin{eqnarray} }
\newcommand {\eea} {\nonumber \end{eqnarray}}
\newcommand {\eps} {\epsilon}

\newcommand {\LL}{{\cal L}}

\newcommand {\lan} {\langle}
\newcommand {\ran} {\rangle}

\newcommand {\cD}  {{\cal D}}
\newcommand {\cF}  {{\cal F}}

\newcommand {\cL}  {{\cal L}}

\newcommand {\cP}  {{\cal P}}

\newcommand {\bc} {\begin{center}}
\newcommand {\ec} {\end{center}}
\newcommand {\bd}{\begin{displaymath}}
\newcommand {\ed}{\end{displaymath}}

\newcommand {\Tr} {\mbox{Tr}}
\newcommand {\sign} {\mbox{sign}}

\def \form#1 {eq. (\ref{#1}) }
\def \parziale#1#2  {{\partial {#1} \over \partial {#2}}}
\def \bi#1 {\typeout{#1} \item}
\newcommand{\Cred}{}
\newcommand{\Cblu}{}

\newcommand{\CgreA}{}

\begin{document}

\title{The mean field theory of spin glasses: 
the heuristic replica approach and recent rigorous results}
\author{ Giorgio Parisi\\
Dipartimento di Fisica, Sezione INFN, SMC  of INFM-CNR,\\
Universit\`a di Roma ``La Sapienza'', \\
Piazzale Aldo Moro 2,
00185 Rome, Italy\\ }

\maketitle

\begin{abstract}

\noindent  The mathematically correct computation of the spin glasses free energy in the infinite range limit crowns 25 years of mathematic efforts in solving this model. The exact solution of the model was found many years ago by using a heuristic approach; the results coming from the heuristic approach were crucial in deriving the mathematical results. The mathematical tools used in the rigorous approach are quite different from those of the heuristic approach. In this note we will review the heuristic approach to spin glasses in the light of the rigorous results; we will also discuss some conjectures that may be useful to derive the solution of the model in an alternative way.

\end{abstract}
\vskip.5cm
{\bf Keywords}
\vskip.3cm
11E95   $p$-adic theory 

46N30  Applications in probability theory and statistics 

60F10   Large deviations 

82B44   Disordered systems (random Ising models, random Schr\"odinger operators, etc.) 

\vskip.9cm

\section{Spin glasses}

Spin glasses  are interesting for many physical reasons; moreover they are the prototype of a complex system,  i.e.  a system with many phases.
Here we will consider only spin glasses with an infinite range interaction, where the mean field approximation is correct (more of less by definition).
In this case \cite{mpv,LH2,CC,LH7, parisibook2} (i.e. the Sherrington Kirkpatrick model),
the Hamiltonian is given by
	\be
	H_{J}[\sigma]=\frac12 \sum_{i,k=1,N}J_{i,k}\sigma_{i}\sigma_{k} \, ,
	\ee
	where the $N$ variables  $\sigma$ are Ising spins, i.e. $\sigma_{i}=\pm 1$.  The $J$ are random variables: they are Gaussian distributed with zero average:
	
	\be 
	E[J_{i,k}]=0 \, , \ \ \ \ E[J_{i,k}^{2}]= N^{-1} \, ,
	\ee
where $E[\cdot]$ denotes the expectation value respect to the random couplings $J$.

For each instance of the system (i.e. a choice of the couplings $J$) we can compute the ground state energy density that is  defined as	
	\be 
	\Cred{ e^{N}_{J}={\min_{\sigma}H_{J}[\sigma] \over N}}\ .
\ee
The normalization factors have been chosen in such a way that  the energy density is a number of order 1, when $N \to \infty$.
In the same way for each choice of the variables $J$ we can define the free energy density $f^{N}_{J}(\beta)$ and the partition function $Z^N_{J}(\beta)$ as 
\be	
 f^{N}_{J}(\beta)= -{\log(Z^N_{J}(\beta)) \over    \beta  N}\, ,  \ \ \ \ Z^N_{J}(\beta)=\sum_{\sigma}\exp(-\beta H_{J}[\sigma])\ .
	\ee

When $N\to \infty$, it can be proved that the fluctuations in the energy density and in the free energy density go to zero with $N$:
	
	\be
	E[(e^{N}_{J})^{2}]- \left(E[e^{N}_{J}]\right)^{2} \to 0 \, , \ \  E[(f^{N}_{J})^{2}]- \left(E[f^{N}_{J}]\right)^{2} \to 0  \ .
	\ee
	Therefore, with probability $1$ when $N$ goes to infinity, all systems have the same free energy density.

	We want to compute
	\be
	\tilde f(\beta)=\lim_{N\to \infty}E[f^{N}_{J}(\beta)] \  ,\ \ \ \ \  \tilde e=\lim_{N\to \infty}E[e^{N}_{J}] \ .	\ee
In the limit $N \to \infty$ the probability distribution of $f_J(\beta)$  becomes a delta function and the average value becomes equal to the most likely value.

The limit $N\to \infty$ and $\beta \to \infty$ can be exchanged\footnote{As suggested by a referee, the exchange of limits is permitted by the elementary inequality $|E[f^N_J (\beta)] - E[e_N^J]| \le \ln( 2)/\beta$.} and therefore we have that

\be
\tilde e= \lim_{\beta \to \infty} \tilde f(\beta) \ .
\ee

The mean values are not the only interesting quantities. We would like to control the whole probability distribution of the free energy ($f$)  at fixed and  large  $N$  (i.e. $P_N(f)$ \footnote{From time to time, when this does not lead to ambiguities  we will not indicate in an explicit way the $\beta$ dependence of the various quantity: the correct, but heavier, notation would be $P_N(f,\beta)$.}).  We can consider various problems:
\begin{itemize} \item The behaviour of the typical fluctuations around the average value. For example it has been conjectured that in the region where the probability is concentrated  we have \cite{Kon1,PARI2}:
\be
P_{N}(f)\approx N^{-5/6} p\left(N^{6/5}(f-\tilde f)\right) \label{BULK} \ ,
\ee
where $p$ is an unknown function and the difference between the r.h.s and the l.h.s. is negligible for large $N$ (unfortunately we do not know how to compute heuristically the function $p$).
\item
Large deviations. For values of the free energy that is {\em smaller} than the most likely value (i.e. $f<\tilde f$), the probability is exponential small in $N$ \cite{Kon1,TALA0}; the following limit exists (for not too large $f$):
\be
L(f)=\lim_{N\to \infty}{-\log(P_N(f))\over N} \ .
\ee

\item Very large deviations. For values of the free energy that is {\em bigger} than the most likely value (i.e. $f>\tilde f$), it has been conjectured \cite{DFM,TALA0} that the probability  decrease faster than any exponential of $N$ and that it is  exponential small in $N^2$  \cite{PARI2}. The following limit should exist (for not too large $f$) and it should be  non-trivial:
\be
\cL(f)=\lim_{N\to \infty}{-\log(P_N(f))\over N^{2}}\ .
\ee
\end{itemize}
Similar questions are well posed also for the ground state energy $e$.

As we shall see below, there are rigorous explicit expressions for $\tilde f$ \cite {TALA0} and for the large deviations function $L(f)$ \cite{PARI1} . On the contrary for the very large deviations function $\LL(f)$, there are only heuristic evaluations \cite{PARI2}. Nothing analytic is know  (only numerical evaluations) on the function $p$ that enters in the scaling form of the probability $P_{N}(f)$ (eq. (\ref{BULK})). The form in eq. (\ref{BULK}) has been conjectured in order to match with the known behaviour of the large deviation function, i.e.
\be
L(f)\propto (\tilde f-f)^{6/5}
\ee
for $f$ near (and smaller than) $\tilde f$ and with the supposed behavior of the very large deviation function, i.e.
\be
\cL(f)\propto (\tilde f-f)^{12/5} \ .
\ee

The reader should notice that the same kind of questions have a known answer for the ground state energy  in the \emph{linear} case where the variable $\sigma$ are continuous variables:

\be
e^{N}_{J}={\min_{\sigma}H_{J}[\sigma] \over N}  \, , \ \ \ \Cblu{\sum_{i=1,N}\sigma_{i}^{2}=N}\ . 
\ee
Indeed in this case the ground state coincides with the largest eigenvalue of a Gaussian random matrix where nearly everything is known \cite{TW,DM}.
Here we are interested in the more difficult case,
\be
\ \ \ \ \Cred{\sigma_{i}=\pm 1} \ ,
\ee
where $e^{N}_{J}$ is {\em not} a function of only the eigenvalues of  $J$.

\section{Some heuristic considerations}
Let us start with a simple theorem \cite{PS} that gives interesting information on what we should {\em not} assume.
For a given system we introduce the connected correlation functions at temperature $\beta$
\be
\ C_{i,k}=\lan \sigma_{i} \sigma_{k} \ran_{Gibbs} -\lan \sigma_{i}\ran_{Gibbs} \lan \sigma_{k} \ran_{Gibbs} \ ,
\ee
where the bracket denotes the average with respect to the Gibbs distribution\footnote{In order to lighten the notation I have not indicated the dependence of both the Gibbs expectation values ($\lan \cdot \ran$) and of the $P[\sigma]$ on $\beta$ and $J$.}:
\be
\lan A[\sigma] \ran _{Gibbs}= \lim_{N\to \infty}\sum_{\sigma}P[\sigma] A[\sigma]\, , \ \ \ \ P[\sigma]={\exp(-\beta H_J[\sigma]) \over \sum_{\sigma}\exp(-\beta H_J[\sigma])}\ .
\ee

It would be  natural to assume that the connected correlations go to zero when $N \to \infty$, more precisely to assume that
\be
\CgreA{\lim_{N\to\infty}{\sum_{i,k} C_{i,k}^2 \over N^{2}}=0} \label{ASS} \ .
\ee
 If the previous assumption \footnote{The previous assumption should be formulated more carefully in order to take into account the  invariance of the Gibbs measure when we change sign to all the spin simultaneously; I will not discuss here this technical detail.} would be valid,  we could derive  a simple  formula for $f(\beta)$ that implies that 
 \be
 e =\lim_{\beta\to \infty} f(\beta)=- \sqrt{{2\over \pi}}\ .
 \ee
The results is reasonable, but in definite disagreement with numerical estimates for $N$ up to $O(10^3)$. However the situation is much worse: the same argument would imply that the entropy $s(\beta)$ satisfies the relation
 \be
\lim_{\beta\to \infty} s(\beta)= -{1\over (2 \pi)} <0 \ .
 \ee
The entropy $s(\beta)$ is non-negative  by definition. We can thus conclude {\em  by absurdum} that 
 \be
\CgreA{\lim_{N\to\infty}{\sum_{i,k} C_{i,k}^2 \over N^{2}}\ne 0} \label{NE} \ .
\ee

In order to understand the importance of the result (\ref{NE}), it is convenient to recall some know facts on clustering states \cite{KASTLE, KASROB, RUELLE}.
Let us consider an infinite translational invariant (non-random) system. In this case one can introduce the general concept of an equilibrium state, i.e. a state that satisfies local equilibrium conditions (e.g.  the DLR equations \cite{RUELLE}). If the Gibbs probability distribution has a limit when the volume becomes infinite, the Gibbs probability distribution is an example of equilibrium state, but it may be not the {\sl unique} equilibrium state.

Now the set of equilibrium states is convex; by the Alaoglu theorem any state can be written as the linear sum of the extremal (pure) states. Therefore we can write
\be
\Cred{\lan \cdot \ran_{Gibbs} =\sum_\alpha w_\alpha \lan \cdot \ran_\alpha}\, , \ \ \ \ \ \ \ \sum_\alpha w_\alpha=1\, ,
\ee
where the sum runs over the pure states (for simplicity I am assuming that the relevant set of pure states is discrete).
A well-known theorem states that in a pure state connected correlation functions vanish at infinity (at large distances), i.e. pure states are clustering states, and vice versa. Intensive quantities (i.e. average on the whole system of local quantities) do not fluctuate in pure states.

The ferromagnetic Ising at zero field at low temperature provides a well-known case: there are two phases (+ and -):
\be
\lan \sigma_{i}\ran _{\pm}=\pm m\ ,\ \ \ \ \ \lan \sigma_{i} \sigma_{k}\ran _{\pm}\approx m^2 \ , \ \ \ \ \  w_+=w_-= \frac12 \, ,
\ \ \  \lan\cdot \ran_{Gibbs}= w_+\lan\cdot \ran_+ + w_-\lan\cdot \ran_- \, ,
\ee
where, for $i$ and $k$ far away,
\be
 \lan \sigma_{i} \sigma_{k}\ran_{Gibbs}  \approx \lan \sigma_{i}\ran_{Gibbs} \lan \sigma_{k}\ran_{Gibbs}\ = m^{2} \ ;    \ \ \  
\lan \sigma_{i} \sigma_{k}\ran_{+} - \lan \sigma_{i}\ran_{+} \lan \sigma_{k}\ran_{+} \approx 0\ . \nonumber
\ee

The situation is more complex for disordered systems, where the Gibbs measure may not have a limit when the volume goes to infinity.
Although the appropriate and sophisticated tools have been forged to deal with this situation, it may be convenient to consider what happens in large, but finite systems, without putting the $\delta$'s and $\eps$'s needed to make the statements sharp\footnote{As a physicist I would also remark that all  the experimental systems are finite, and a statement on the behaviour of large finite systems can be experimentally tested much better than a statement on the behaviour of an infinite systems.}.

Let us consider a system where the number of degrees of freedom ($N$) is large, but finite. We suppose \cite{PAR1,CINQUE} that for large $N$ we can write
\be
\lan \cdot \ran_{Gibbs} \approx \sum_\alpha w_\alpha \lan \cdot \ran_\alpha\ ,  \ \ \ \ \ \ \ \sum_\alpha w_\alpha \approx 1 \ . \label{GIBBS}
\ee
The states are approximately clustering if 
\be
{\sum_{i,k} (C^{\alpha}_{i,k})^2 \over N^{2}}= o(1) \ .
\ee

The previous theorem (see eq. (\ref{NE})) tells us that for spin glasses more than one $w$ should be different from zero \footnote{More precisely more than  two $w$'s.} and therefore the decomposition in eq. (\ref{GIBBS})  is non-trivial.  In order to describe better the Gibbs decomposition  is convenient to introduce additional concepts.

The first interesting quantity is the probability distribution of the $w$'s (i.e. $\cP(\{w\})$); however  the weights $w$'s of the states do not convey the whole information. We would like to know how  the states differ one from the others: at this end we can introduce the distance and the overlap between two states:

\be d_{\alpha,\gamma}= {\sum_{i=1,N}\left( \lan \sigma(i) \ran_\alpha - \lan \sigma(i) \ran_\gamma\right)^{2}\over N }\ ; 
\ \ \ \ q_{\alpha,\gamma}= {\sum_{i=1,N}\lan \sigma(i) \ran_\alpha \lan \sigma(i) \ran_\gamma\over N}\ .
\ee

Obviously the distance $d$ and the overlap $q$ are related. We have 
\be
d_{\alpha,\gamma}=2(q_{EA}-q_{\alpha,\gamma}) \, ,
\ee
where $q_{EA}\equiv q_{\alpha,\alpha}=q_{\gamma,\gamma}$ (we are assuming that $q_{\alpha,\alpha}$ is independent form $\alpha$). Usually in literature one studies the properties of the overlap, the distances have been introduced here because their definition is more natural.

 \Cred{Now for a given system $J$, we introduce the  descriptor $\cD_J$, that is  given by the following set} 
 \be \{  w_{\alpha}, q_{\alpha,\gamma} \} \ , \ee
 where the index $\alpha$ runs over a numerable set\footnote{The numbers of states is obviously finite for finite $N$, however in order to consider the limit $N \to \infty$ it is useful to consider a numerable set of states.}.
 
The descriptor $\cD_J$ does depend on $J$. The functional   $\cP(\cD)$ is the probability distribution of the  descriptors. One can construct heuristic arguments  \cite{mpv} to compute the probability distribution  $\cP(\cD)$, the free energy $\tilde f$ and the large deviation function $L(f)$.
We will describe these results later on in section (4). We now proceed into the description of some rigorous results.

\section{Rigorous results}

In this section we will concentrate our attention on some theorems. They are obtained using two different approaches, the first is based on probabilistic techniques, the second heavily relies on Guerra interpolation technique.

Without any reference to the physical meaning of the descriptor and to the Gibbs phase decomposition we can  introduce  the descriptor $\cD$  that is defined as a normalized vector and a matrix, satisfying certain inequalities \cite{V2}. Neglecting technical details, a descriptor ($\cD$) is given by the set of $w_\alpha$ and of $q_{\alpha,\gamma}$, where $\alpha$ belong to a numerable set and $\sum_\alpha w_\alpha =1$. 
In this context one would naturally associate to a descriptor a decomposition of the Gibbs measure into states, however for the sake of proving theorems one can define a descriptor in an abstract way. Moreover we can also introduce a probability distribution on the space of the descriptors ($\cP(\cD)$). The following theorems hold.

\begin{itemize}
\item One can define  a free energy functional $\cF[\cP]$ ($\cP(\cD)$ being the probability of descriptors) such \cite{V2} that the true free energy is given by
\be
\tilde f(\beta) = \max_\cP \cF[\cP] \ .
\ee
The explicit form of the free energy functional $\cF[\cP]$ is rather complex  \cite{V2} and it will be not described here.
\item 
Using arguments, which do not refer to descriptors and are based on Guerra interpolation, it is possible to compute in an explicit way the value of  $\tilde f$ and consequently (from the previous theorem) $\max_\cP \cF[\cP]$.
By inspection one verifies that the maximum is reached for the probability distribution predicted by the heuristic approach  ($\cP^{*}$): also the value of the free energy $\tilde f(\beta)$ coincides with the one computed by the heuristic approach \cite{V1,TALA}. Therefore we know a probability distribution that maximize $\cF[\cP]$.
\item Under  some assumptions, i.e. if stochastic stability and ultrametricity (see the definitions later) hold \cite{PRzz}, the function 
\be
P(q)=\overline{\sum_{\alpha,\gamma}w_{\alpha}w_{\gamma}\delta(q-q_{\alpha,\gamma})} \label{PQ}
\ee
 determines the probability distribution  $\cP(\cD)$. Therefore in the stochastic stable ultrametric case the 
probability distribution is essentially unique, apart from a redefinition of the overlap.
\end{itemize}

The bad  new is that we do not know  if the true $\cP(\cD)$ has some relation with the $\cP^{*}(\cD)$ that maximizes the free energy; for the moment the probability distribution $\cP(\cD)$ does not have a definite meaning. It also not clear if the maximum of the free energy functional  is unique, also if we consider only those probability that are stochastically stable \footnote{The reader can find the definition of stochastic stability and a discussion on its consequences in \cite{GUERRA,SS,SS1,SS2}.}.

It may be interesting to note the maximum is reached for the probability an ultrametric $\cP^{*}$:  i.e. the probability matrix is concentrated on distances that satisfy the ultrametric inequalities: 
\be
d_{\alpha,\gamma}\le \max (d_{\alpha,\beta},d_{\beta,\gamma}) \ \ \ \forall \beta \ .
\ee

Ultrametricity implies that the probability distribution of the distance among three random configurations  is supported, in the limit of very large systems, only on equilateral and isosceles
triangles with no contributions coming from  scalene triangles. In an ultrametric space if two spheres have one point in common, their union coincides with the sphere having maximum radius: in other words a random walk with steps of length $l$ would never arrive at a distance larger than $l$ from the origin.   \begin{figure} \begin{center}    
\includegraphics[width=0.65\textwidth]{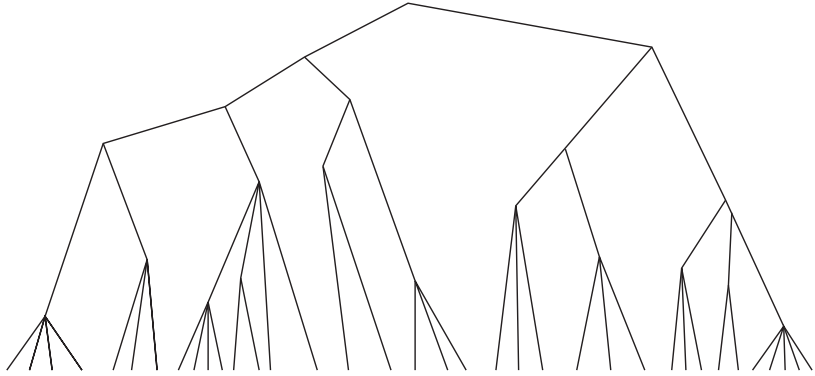}
      \end{center} \caption{ \label{ULTRA} A taxonomic tree: the leaves denote the states: they are a the bottom. The distance among the states corresponds to the minimum height we have to cross for going from a state to an other state. This definition of the distance is ultrametric. }
 \end{figure}  

Ultrametricity is a very striking property: it implies that the equilibrium configurations of a large system
can be classified in a taxonomic (hierarchical) way (as animals in different taxa).
When ultrametricity holds, the space of the descriptors reduces to the space of infinite weighted trees (each leave has a weight). Ruelle gives a precise mathematical definition of a probability distribution over these trees  \cite{RUELLETREE, PT,FB} (in his paper  one can also find interesting comments on the pruning of infinite trees.)

\subsection{Some details on the solution}
In this section we present the explicit form for the free energy $\tilde f(\beta)$. 
We firstly introduce an non-decreasing function $q(x)$ that is defined in the interval $[0-1]$. In the replica approach the inverse function $q[x]$ is related to the function $P(q)$ (defined in eq. (\ref{PQ})) by the equation: 

\be
x(q)=\int_0^q d \,q'P(q')\ .
 \ee

Let us consider a  function $g(x,h)$, that  is defined in the strip $0\le x\le 1$.  The function $g$ satisfies the following boundary condition at $x=1$:
\be
g(1,h)=\log(\cosh(\beta h))\ .
\ee
Moreover (if we assume for simplicity that $q(x)$ is differentiable), the function $g(x,h)$ must be the unique solution of the following antiparabolic equation

\be{\partial g\over \partial x}=-{dq\over dx}
\left({\partial^{2} g\over \partial h^{2}}+x\left({\partial g\over \partial h}\right)^{2}\right) \ . 
\ee
We now introduce a functional $F[q]$:
\be
F[q]=\beta \int_{0}^{1} dx\left(1- q(x)^{2}\right) -g(0,0) \ . \label{FIN}
\ee
Talagrand theorem tell us that the free energy $F$ is given by
\be
F=\max_{q(x)}F[q] \ .
\ee
Moreover, using Ruelle construction \cite{RUELLETREE}, we can associate to each function $q(x)$ a probability distribution over the descriptors, i.e. $\cP(\cD)$. For  the probability distribution over the descriptors  $\cP(\cD)$  that corresponds to a function $q(x)$, the functional $\cF[\cP]$ (introduced in  \cite{V2}) gives the same result of eq. (\ref{FIN}).

These results extend to large deviations. Indeed a theorem of Talagrand states that for $1>n>0$ we can define in the large $N$ limit a large deviation function $\hat f(n)$:
\be
\int dP_N(f) \exp (-N n \beta f) \approx \exp( - N\beta \hat f(n)) \ .
\ee
The functions $\hat f(n)$ and $L(f)$ are simple related by a Legendre transform at usual.
Talagrand has proved that the explicit value of $\hat f(n)$ is given by
\be
\hat f(n)= \max _{q(x)}\left[\beta\int_{n}^{1} dx \left(1-q(x)^{2}\right) -g(n,0)\right]  \ \ \ \ n<1 \ \ \ \ \ \  q(x) \ \mbox{for} \ n\le x\le 1 \ .
\ee
It is evident that for $n=0$ we have that $\hat f(0)=\tilde f$, as it should be.

It is possible to perform analytic and numerical evaluations of the function $\hat f(n)$ following \cite{Kon1,PARI1}, where the properties of the function $\hat f(n)$ are discussed in details.

When  $n<0$, there are no theorems. 
There is a conjecture \cite{DFM}:
\be
\lim_{N\to\infty}{\log\left(\int dP_N(f) \exp (-N n \beta f)\right)\over N n \beta }= \hat f(0)\equiv \tilde f \ .
\ee
In other words the probability $P_N(f)$ should be smaller than an exponential of $N$ if $f>\tilde f$.

What about very large deviations? In \cite{PARI2} it has been conjectured that
\be
\int dP_N(f) \exp \left(-N^{2}n \beta (f-\tilde f)\right) \approx \exp\left( - N^{2}\beta \hat{f}_{LL}(n)\right) \ .
\ee
Moreover some heuristic computations of the function $\hat{f}_{LL}(n)$ have been performed and one arrives the following formula:
\be
P_N (f) \approx \exp \left(-C N^{2} |f-\tilde f|^{12/5}+\ldots\right)\, ,
\ee
where the value of $C$ is approximately evaluated \cite{PARI2} and it should be near to 2/3.

\section{The algebraic replica approach}

In the original  heuristic approach the computations were done using the algebraic replica method \cite{mpv}. Later it was found that the replica method was equivalent to a probabilistic approach \cite{mpv} that inspired the rigorous results \cite{V1,V2,TALA}. However it may worthwhile to try to understand the basis of the replica method in order to see if we can  give it a rigorous basis. This is not easy because the algebraic replica approach is based on very strange mathematics.
	
	\Cblu{The mean field theory was solved by replica approach.} The basic idea is quite simple and  essentially it goes back to Nicola d'Oresme.
Our goal is  to compute  \be
	\tilde f(\beta) \equiv f_\infty(\beta)= \lim_{N\to\infty}f_N(\beta)\, ,\ \ \    f_N(\beta)=- { E\left[\ln \left(Z_N^J(\beta)\right)\right] \over \beta N}\ .
	 \ee
	However in our case  it is much simpler  to compute  (for integer $n$)
		\be
	f_N(n,\beta)=- {\log \left(E\left[\left(Z_N^J(\beta)\right)^n\right]\right)\over\beta N n} \, ,
	\ee
	and at the end performing the limit
	\be
	f_\infty(n,\beta)=\lim_{N\to\infty}f_N(n,\beta).
	\ee
	Although the computations are simple for integer $n$ the construction can be done also for non-integer  $n$ and we have that 
	\be
	\lim_{n \to 0} f(n,\beta)=f_\infty(\beta) \  .
	\ee
	The function $f_\infty(n,\beta)$ is interesting because it is  the one relevant in the large deviations regime. 

	A simple computations can be done for integer $n$, while we need the function $f_\infty(n,\beta)$ also for non-integer $n$. While $f_N(n,\beta)$ is an analytic function $n$, at low temperatures (high $\beta$) $f_\infty(n,\beta)$ is not an analytic function of $n$ so that we cannot obtain its value at $n=0$ by analytic  continuation in $n$ from integer $n$. 

In order to bypass this problem physicists were lead to an unusual mathematics: as we shall see below one introduces an $n \times  n$ matrix $Q$, where eventually $n$ is analytically continued to 0.
	 
	 An elementary computation (based on Gaussian integrals) tell us that when $n$ is integer
	 \be
	 \exp (-\beta n N f_N(n,\beta) )= \log \left(E\left[\left(Z_N^J(\beta)\right)^n\right]\right)=\int d Q \exp(-\beta N n F_\beta[Q]) \label{INTE} \, ,
	 \ee
	 where $Q$ are $n \times n$ symmetric matrices (that are zero on the diagonal) and the integral is done over all these matrices (the function $F[Q]$ has a simple expression).
	 
	 It is evident that for positive integer $n$ the integrals in equation (\ref{INTE}) can be evaluated by the saddle point method and one gets.  
	\be
	 f_\infty(n,\beta) =\min_{Q}F_\beta[Q])=F_\beta[Q^*] \ .
	 \ee
Unfortunately, at high $\beta$  the function $ f_\infty(n)$ that is defined for all $n$, is  non-analytic in $n$. So in order to evaluate it in the interesting region and to find 
out the value of $Q^*$ one has to consider integral representation for non-integer $n$. In this way one is lead to consider the saddle point over $0 \times 0$ matrices. 

There are many ways in which a $0 \times 0$ matrix may be constructed.  At the end of the day in the replica approach the matrix $Q$ is parameterized  in terms of a function $q(x)$ defined on the interval $0-1$, so that the space of $0 \times 0$ matrices becomes infinite dimensional. The  game consists in computing everything by analytic continuation of the integral representation.

There are many way equivalent to do the construction and they lead to the same result. We choose one that is particularly simple \cite{PASOU,PADIC}.

Let us take \Cred{a prime number $p$} and let us assume that there is an integer $m$ such that $p^{m}=n$.
We look for saddle point of $F_\beta[Q]$, where the matrix $Q$ has the form:
\be
Q_{a,b}=q\left(|a-b|_{p}^{-1}\right)\ .
\ee
Here $|a|_{p}$ is the $p$-adic norm: (the $p$-adic norm is defined as follows: if $p^{k}$ divides $a$ and $p^{k+1}$ does not divides $a$, $|a|_{p}=p^{-k}$).

We write $F_\beta[Q]$ as function of $p$ and  $m$ we make an analytic continuation in $p$ and $m$ of the result. {\em Everything} is computed by making analytic continuations from the region where the construction make sense. We will also write down the equations for the critical points in an explicit way for the function $q(x)$. The final recipe consists in assuming that the free energy is given by the analytic continuation of the free energy that is evaluated at the analytic continuation of the saddle point. There are no serious mathematical justifications for this procedure. The heuristic justifications are:
\begin{itemize}
\item At the present moment in the framework of the replica approach there is nothing else that you can try. You have two choices: or you accept this approach, or you change framework (or you invent something new).
\item It works, it gives results that have sense and that have been confirmed by a rigorous mathematical analysis.
\item It is quite likely that the rigorous results would have not been derived without the heuristic analysis of the replica method.
\end{itemize}

Now we can proceed to the limit $n \to 0$: at this point nothing is anymore an integer. Let us suppose that the {\em prime number} $p$ satisfies the condition $0<p<1$; this is quite possible  because $p$ is not anymore a prime number, but it is the analytic continuation of a prime number. Now if we send $m \to \infty$ the quantity $n$ goes to zero, as it should be ($p$ is less than 1).

As a final step we perform the limit  \Cred{$p \to 1^{-}$}: the range of $|a-b|_{p}^{-1}$ becomes the interval[ $0-1]$ and the matrix $Q$ depends on the function  $q(x)$ defined on the interval $[0-1]$.
In this way the function $F[Q]$ is promoted to a functional of $q(x)$ and one finds the previous formulae for $F[q]$. The replica approach  gives the exact result for the free energy and the functions $q(x)$ is the same of the one of the mathematical approach.
It is remarkable that integers are an ultrametric space with respect to the $p$-adic distance and this ultrametricity property implies the ultrametricity of the distances in the descriptor.

Later one it was  a surprise to discover was that all the formulas of the replica approach can be translated into probabilistic statements \cite{mpv}.
The matrix $Q$ of the replica approach is a very compact to code the probability $\cP(\cD)$. 
It looks strange that a matrix codes a probability distribution, so that let us explain how it works. 

One can prove the following formula
\be
E\left[\delta(Q_{a,b}-q)\right]= E\left[\sum_{\alpha,\beta} w_\alpha w_\beta \delta(q_{\alpha,\beta}-q)\right]\,, \label{MAGIC}
\ee
where the $E$ at the l.h.s. denotes the average over the indices $a$ and $b$ (with $a \ne b$) and $E$ at the r.h.s. denotes the average over the probability distribution of the descriptors. In a similar way one can prove
\bea
E\left[\delta(Q_{a,b}-q_{1,2})\delta(Q_{b,c}-q_{2,3})\delta(Q_{c,a}-q_{3,1})\right]=\\
 E\left[\sum_{\alpha,\beta,\gamma} w_\alpha w_\beta w_\gamma\delta(q_{\alpha, \beta}-q_{1,2})\delta(q_{\beta, \gamma}-q_{2,3})\delta(q_{\gamma,\alpha}-q_{3,1})\right]\ .
\eea
Going on one can write an infinite set of relations that should uniquely determine the probability distribution over the descriptors.
It is quite remarkable that a $0\times 0$ matrix is used to code the probability distribution over a space of infinite dimensional matrices and that  the ultrametricity of the integers with respect to the $p$-adic norm implies the ultrametricity of the matrix of distances among states.

The reader should note that not all $Q$ matrices codes for a probability: the probability must a positive function and there are many matrices Q where the l.h.s. of eq. (\ref {MAGIC}) cannot be interpreted as a probability. It is also not clear if there are probabilities  $\cP(\cD)$ that cannot be coded using the matrix $Q$: a negative result that is difficult to prove and there is no constructive procedure to get the matrix $Q$ (if any) from the probability distribution of the descriptors.

In conclusions the heuristic version of the descriptor  approach (i.e. the cavity method) and the replica approach are equivalent \cite{PAR1}.
\Cred{The functions $q(x)$ of both approaches  are equal.} \Cblu{The replica approach  gives the exact result for the free energy.}

The reader may be perplexed by this procedure that was invented about thirty years ago. However it has a very strong heuristic value and both the heuristic probabilistic approach and the rigorous results have strongly beneficed by the fact that the answer to many question was known especially in this context that is rather unfamiliar.

\section{Two conjectures}
Can the original replica derivation made rigorous? The distance from conventional mathematics seems to be very far so that a direct assault  may be not successful. However it is clear that the subject of our discussions is the set of critical points of the function $F[Q]$ for all $n$, in the same way as the analytic continuation of a function on the integers is related to the value that the function takes on all integers.

Following \cite{CPV} we would like to  present some conjectures, whose proof would be a step forward a rigorous understanding of the replica approach.

Let us consider a function $Z(n,N)$ that can be simultaneously written as 
\be
\Cred{ Z(n,N)=\int_{0}^{+\infty} dz \mu_{N}(z) z^{n}=\int dQ \exp(-N\, F(Q))} \, ,
\ee
where  $\mu(z)$ is a positive function and the integral runs over symmetric $n \times n$ matrices; $F(Q)$ is an analytic function of $n$, in the sense that it has a fixed form for all $n$ \footnote{Examples of allowed functions are $\Tr (Q^4)$ or $\sum_{a,b} Q^4_{a,b}/n$, the factor $1/n$ is needed in order to have a non zero limit at $n=0$ in interesting cases.}. The existence of functions, which can be written in both ways, is non-trivial, however the partitions function of spin glasses and generalized spin glasses provide an explicit example.

We know define a new function $\tilde Z(n,N)$, i.e.

\be
\tilde Z(n,N)=\sum_{k} \exp\left(-N\, F(Q_k)\right)
\ee
and the sum goes over all the critical points of the function $F$, i.e. those satisfying the condition
\be
{\partial F \over\partial Q}\Big|_{Q_{k}}=0\, .\ee

An alternative definition  could be
$
\tilde Z(n,N)=\sum_{k} s_k \exp (-N\, F(Q_k))
$,
where  the factors $s_k$ are given by
$
s_k= \sign\left(\det\left({\partial^{2} F \over\partial Q^{2}}\right)\right)\Big|_{Q_{k}}
$.
It not clear which of the two definitions is the most appropriate.

Our aim would by to approximate $Z(n,N)$ with $\tilde Z(n,N)$ also for non-integer $n$. At this end we need to have some {\em a priori} control on the analytic properties of both functions. In this context we put forward the following conjectures.
\begin{itemize}
\item The function $\tilde Z(n,N)$ satisfies for integer $n$ the following integral representation.
\be
\CgreA{\tilde Z(n,N)=\int_{0}^{+\infty} dz\tilde \mu_{N}(z) z^{n}} \ .
\ee
The previous formula implies the existence of a natural (and under certain conditions unique) analytic continuation of
$\tilde Z(n,N)$.
\item The function $\tilde Z(n,N)$ is a good and uniform approximation to $Z(n,N)$ for large $N$: we should have:
\be
\lim_{N\to \infty}{\log\left(\tilde Z(n,N)\right)\over N}=\lim_{N\to \infty}{\log\left( Z(n,N)\right)\over N}
\ee
The previous formula should
not only for integer positive $n$, where it is trivial, but also for non-integer $n$ and in particular for $n$ in the interval $[0-1]$.
In the best of possible words  one should have: 
\be
\lim_{N\to \infty}{\log\left(\tilde\mu_{N}(x^{N})\right)\over N}=\lim_{N\to \infty}{\log\left(\mu_{N}(x^{N})\right)\over N}\ .
\ee
\end{itemize}
These conjectures are interesting also because not many results are known in this direction (as far as I can tell). However in the contest of this paper they are very relevant because in  some cases these conjectures have been used   to prove the result of coming from the replica approach \cite{CPV}.

\end{document}